\documentclass[aps,prb,twocolumn,groupedaddress]{revtex4-1}
\usepackage{graphicx, mathrsfs, amssymb, amsmath, bm}

\begin{document}

%\preprint{}

\title{Anomalous organic magnetoresistance from competing carrier-spin-dependent interactions with localized electronic and nuclear spins}


\author{Y. Wang$^\dag$, N. J. Harmon$^\dag$, K. Sahin-Tiras, M. Wohlgenannt, and M. E. Flatt\'e}
\email[]{michael\_flatte@mailaps.org}
\affiliation{Department of Physics and Astronomy, Optical Science and Technology Center, University of Iowa, Iowa City, IA 52242}
\affiliation{$\dag$These authors contributed equally to this work.}


\date{\today}

\begin{abstract}
We describe a new regime for low-field magnetoresistance in organic semiconductors, in which the spin-relaxing effects of localized nuclear spins and electronic spins interfere. The regime is studied by the controlled addition of localized electronic spins to a material that exhibits substantial room-temperature magnetoresistance ($\sim 20$\%). Although initially the magnetoresistance is suppressed by the doping, at intermediate doping there is a regime where the magnetoresistance is insensitive to the doping level. For much greater doping concentrations the magnetoresistance is fully suppressed. The behavior is described within a theoretical model describing the effect of carrier spin dynamics on the  current.
\end{abstract}

%\keywords{}

\maketitle


%%\section{Introduction\label{sec:introduction}}

Organic semiconductors (OSC) exhibit  intriguing room-temperature spin-dependent phenomena, including magnetic field effects on ``radical pairs''.\cite{Schulten:1978:JChemPhys, Francis:2004:NewJPhys,Prigodin:2006:SynthMet,Hu:2007:NatMat,Bloom:2007:PRL,Kalinowski:2003:ChemPhysLett,Mermer:2005:PRB, Desai:2007:PRB,Bobbert:2007:PRL}  Radical pairs, which are spin-carrying excitations that occupy neighboring molecules in an organic film, can consist of electron-hole pairs, electron-electron (or hole-hole) pairs, and mixed pairs consisting of a spin-1/2 polaron and a spin-1 triplet exciton. These radical pairs can undergo spin-dependent reactions, which due to the large on-site exchange energies in OSC, depend sensitively on the pair spin state. The spin-dependent behavior of radical pairs that occupy transport bottleneck sites can have a significant effect on the conductivity\cite{Bobbert:2007:PRL, Harmon:2012:PRL} and electroluminescent efficiency\cite{Kalinowski:2003:ChemPhysLett, Macia2014} of an organic device. The spin dynamics of these radical pairs is sensitive to the presence of magnetic fields, including an applied magnetic field, an exchange or dipolar field with neighboring spins (localized or mobile), and a nuclear (hyperfine) field. Low-field, room-temperature magnetoresistance in OSC with nonmagnetic electrodes, so-called ``OMAR'', is one consequence.  Even though the exact mechanism behind OMAR is still debated\cite{Boehme2013}, it is widely believed to be related to hyperfine interactions and radical pair effects as described above, at least in some materials. The effects of additional radical dopant spins on these phenomena were first studied in the context of organic solar cells\cite{Zhang2012, Zhang2013}.
The efficiency and short circuit current of doped  P3HT/PCBM solar cells were maximum for a certain doping percentage ($\sim 3\%$), which was suggested to originate from galvinoxyl spins near the PCBM side of the P3HT/PCBM boundary that interacted with electrons to decrease recombination losses by facilitating intersystem crossings to triplet states.

Here we examine, experimentally and theoretically, the influence of radical doping on the transport characteristics in a single organic semiconductor (MEH-PPV). We find that for initial doping the effect of the radical spins is to relax the spins, thus reducing the effect of magnetic fields on the organic magnetoresistance. For intermediate doping ranges, however, the saturation value of the organic magnetoresistance plateaus. This corresponds to doping densities and interaction strengths for which a dopant spin interacts strongly with only one component of the radical pair. This interaction interferes with the typical transport bottleneck behavior of a radical pair. The consequence is a reduction of the OMAR to half its undoped value; the value of this plateau is  very weakly dependent on temperature and other device characteristics. For further doping a dopant spin interacts strongly with both components of the radical pair, and for sufficiently large doping the OMAR is fully suppressed. We describe this effect theoretically in a model of spin-dependent recombination, including the hyperfine field and the dopant spin. We reproduce quantitatively the value of the OMAR plateau, and qualitatively the doping dependence of OMAR.
Our conclusions also support the tentative interpretations of the mechanism in doped organic solar cells in Refs. \onlinecite{Zhang2012, Zhang2013}.

\begin{figure}[ptbh]
    \begin{centering}
        \includegraphics[scale = 0.25,trim =37 5 16 10, angle = -0,clip]{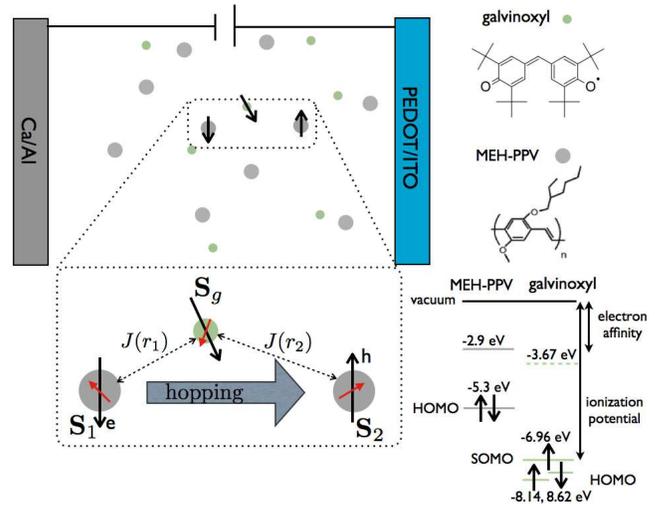}
        \caption[]{(Color online) Schematic of device structure. Dopants (green, smaller symbols) in an organic host material (larger, gray symbols). Bottleneck in host is denoted by sites $1$ and $2$. Black arrows (vertically aligned) are electron or hole spin vectors. Red arrows (randomly oriented) are hyperfine field vectors. Lower right: HOMO, LUMO, and SOMO levels for MEH-PPV and galvinoxyl as well as the electron affinity and ionization potential for galvinxoyl.\cite{Zhang2012, Zhu2004}}\label{fig:diagram}
    \end{centering}
\end{figure}

Galvinoxyl (a commercially available free radical with excellent stability, see Fig.~\ref{fig:diagram} for its chemical structure) was purchased by Aldrich and added at varying concentrations to MEH-PPV solutions in toluene. OMAR devices based on films (thickness of around 130~nm) spin-cast from these solutions were then fabricated (see Fig.~\ref{fig:diagram}). More specifically, the devices' fabrication started with glass substrates coated with 40~nm of indium tin oxide (ITO), which were cleaned by ultrasonic and plasma treatment. ITO, coated with a $\approx$100~nm thick layer of the conducting polymer Poly (3,4-ethylenedioxythiophene)-poly (styrene-sulfonate) (PEDOT:PSS), serves as the anode or hole-injecting contact.  Then the organic galvinoxyl-doped MEH-PPV layer was spin-cast. The top/electron-injecting electrode (cathode) consisting of calcium (20~nm) and a capping layer of aluminum (40~nm) was fabricated by thermal evaporation at a base pressure of $10^{-7}$~mbar. The device area was roughly 1$\times$1 mm$^2$.

The samples were measured inside the dynamic vacuum of a cryostat located between the poles of an electromagnet. All data shown in this manuscript are for room temperature, unless specified otherwise.

\begin{figure}
  \includegraphics[width=0.9\columnwidth]{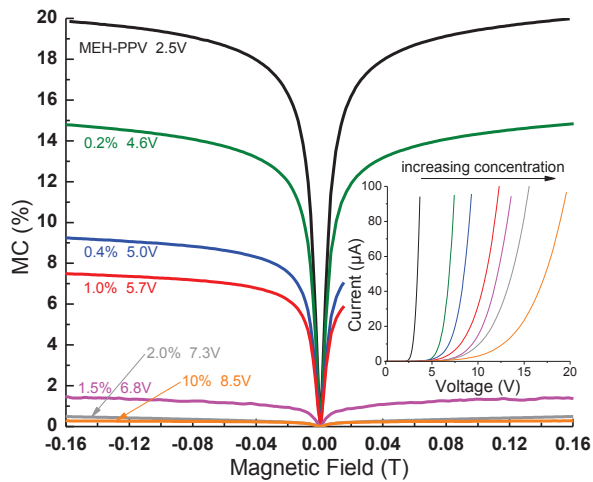}\\
  \caption{Room-temperature magnetoconductance traces for a constant applied voltage for several galvinoxyl doping concentrations. Current flow through sample was approximately 1.5~$\mu$A. Inset: The current-voltage characteristics of the device studied.}\label{fig:OMARtraces}
\end{figure}

Fig.~\ref{fig:OMARtraces} shows the room-temperature magnetoconductance traces for a constant applied voltage for several devices with different concentrations of the radical dopant. The corresponding current-voltage (I-V) characteristics of the devices are shown in the inset. Note that the addition of galvinoxyl does not dope the device in the sense of enhancing conductivity; rather the galvinoxyl radicals form immobile spins, as expected from the location of its energy levels compared to those of MEH-PPV (see Fig.~\ref{fig:diagram} for energy levels). The OMAR trace for the pure MEH-PPV film closely resembles traces commonly measured in a variety of organic semiconductor devices \cite{Mermer:2005:PRB}. The characteristic width of the traces (about 10 mT) corresponds to the hyperfine interaction scale. The OMAR traces for the devices with galvinoxyl dopants exhibit the same line shape, but the magnetoconductance magnitude decreases with increasing radical dopant concentration.

\begin{figure}
  \includegraphics[width=0.9\columnwidth]{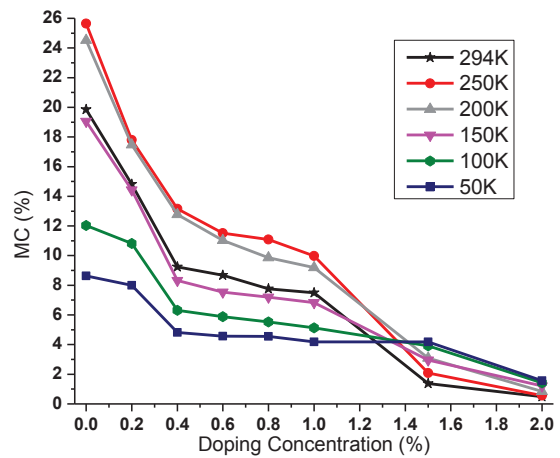}\\
  \caption{Dependence of the magnetoconductance magnitude at 0.1T on the galvinoxyl concentration for several different temperatures. The devices were operated at a constant voltage, resulting in a current flow of approximately 1.5 $\mu$A.}\label{fig:concentration}
\end{figure}

\begin{figure}
  \includegraphics[width=\columnwidth]{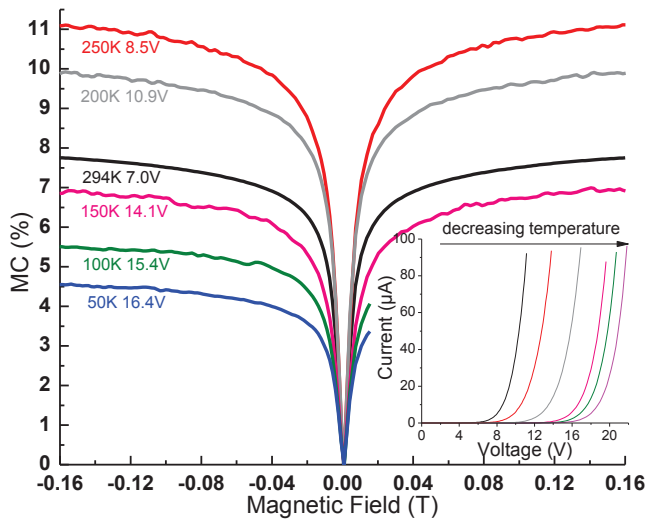}\\
  \caption{Magnetoconductance traces for a device with 0.8\% galvinoxyl concentration for several different temperatures at a current flow of 1.5 $\mu$A. The inset shows the I-V traces for the same temperatures.}\label{fig:temperature}
\end{figure}

The dependence of the OMAR magnitude on the galvinoxyl concentration is shown in more detail in Fig.~\ref{fig:concentration} for several temperatures from room-temperature down to 50 K. This dependence has a characteristic feature that is present at all temperatures. The OMAR magnitude initially decreases rapidly with increasing doping concentration, but reaches a plateau at intermediate doping concentrations, before finally decaying towards zero magnitude at the largest doping concentrations. The theory that will be developed below is able to reproduce this characteristic feature. For completeness, Fig.~\ref{fig:temperature} shows the dependence of the OMAR traces for one sample (0.8\% galvinoxyl doping) for several different temperatures. The inset shows the I-V traces for the same temperatures.

\begin{figure*}[ptbh]
 \begin{centering}
        \includegraphics[scale = 0.425,trim = 5 280 25 110, angle = -0,clip]{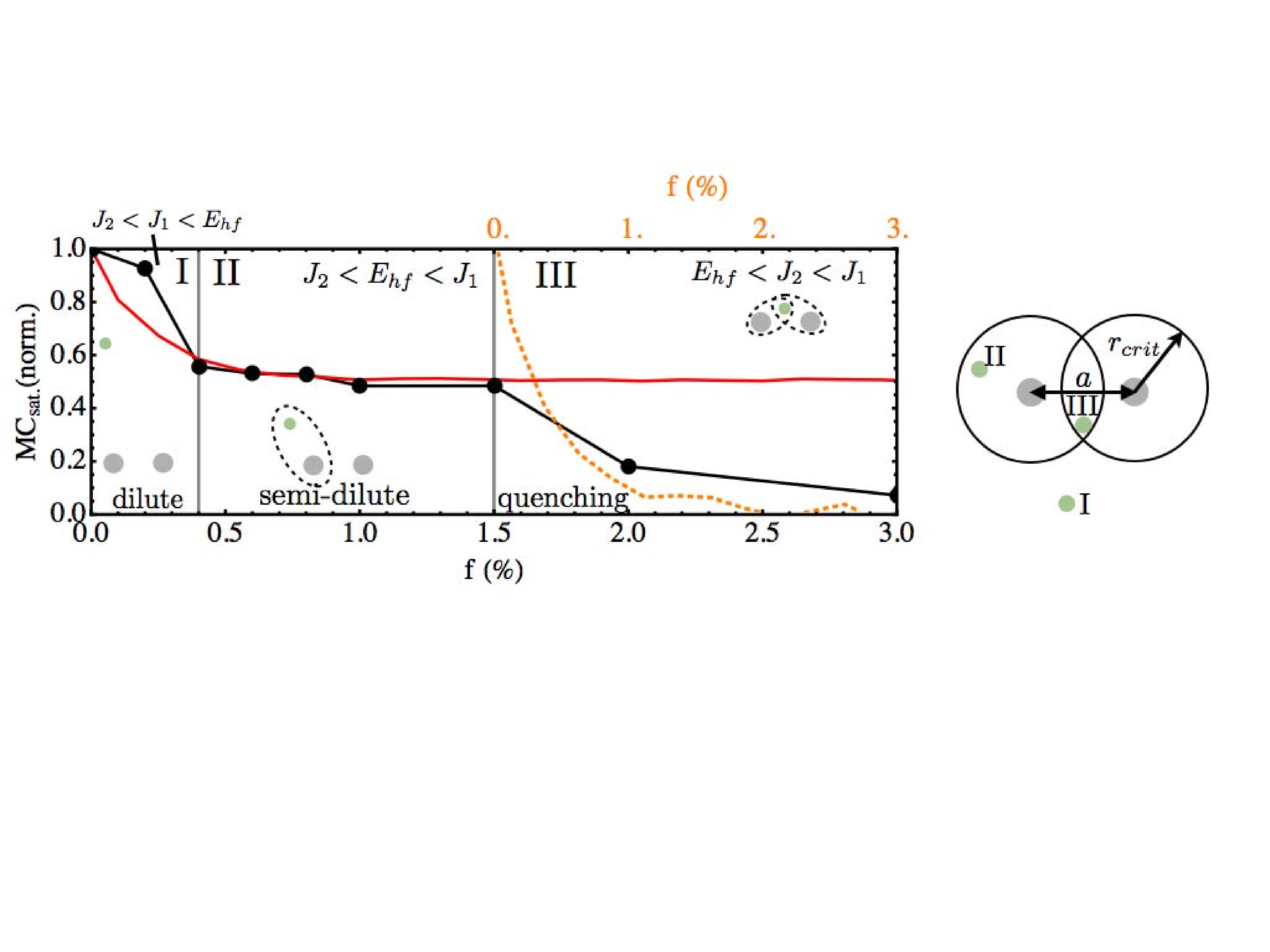}
        \caption[]
{(Color online) Experiment (50 K; black symbols connected with black line for aide to eye) and theory (solid red line and dotted orange line) clearly demonstrating three regimes of doping effects on magneto-transport.
Region I: galvinoxyl molecule is far from bottleneck such that both exchange couplings are smaller than the hyperfine coupling.
Region II: galvinoxyl molecule is now near one of the bottleneck sites such that its exchange coupling with that site is larger than $E_{hf}$ (inequality denoted by encircling dashed line). The exchange coupling with the further site however remains small compared to $E_{hf}$.
Regime III: Exchange couplings between galvinoxyl and bottleneck spins now dominate the hyperfine interaction. The galvinoxyl acts as an intermediary for exchange interaction between the MEH-PPV sites which is known\cite{Wagemans2011, Harmon2012c} to effectively pin the spins thereby quenching the magnetoconductance.
Right: alternative image of the three different regimes.}\label{fig:regimes}
        \end{centering}
\end{figure*}

We now develop a minimal model to explain the MC's doping dependence.
In particular we seek to explain the temperature-independent decay-plateau-decay behavior observed in Figure~\ref{fig:concentration}.
To do so, we describe  three different regimes, displayed in Fig.~\ref{fig:regimes}, and provide qualitative arguments assuming only a \emph{single} galvinoxyl spin interacting with the bottlenecked sites.
This assumption is believed to be justified because the density of galvinoxyl ($n_g$) is small compared to the density of MEH-PPV sites ($n_m$); $n_g  = f n_{m}$ with $n_{m}  = 1.428 \times 10^{21} $ cm$^{-3}$. The mean distance between dopants (host sites) is about $0.893 \lambda^{-1/3}$ where $\lambda = n_{g, m} \pi^{3/2}/\Gamma(5/2)$.\cite{Hertz1909, Torquato1995}
The average inter-distances are then $\langle r_m \rangle \approx 0.5$ nm, $\langle r_g \rangle \approx  3.9$ nm at $f \approx  0.002$, and $1.8$ nm at $f = 0.02$.

Our spin Hamiltonian is given by
\begin{eqnarray}
&&\mathscr{H}' = \mathscr{H}'_{hf} + \mathscr{H}'_{Z} + \mathscr{H}'_{exc}
 = \nonumber\\
 &&{}(\bm{B}'_{{hf}_1} + \bm{B}'_0) \cdot \bm{\sigma}_1 +  (\bm{B}'_{{hf}_2} + \bm{B}'_0) \cdot \bm{\sigma}_2 +(\bm{B}'_{{hf}_g} + \bm{B}'_0) \cdot \bm{\sigma}_g{} \nonumber \\
 &&{} + \frac{1}{2} J'(r_1) (1+ \bm{\sigma}_1 \cdot \bm{\sigma}_i)
  + \frac{1}{2} J'(r_2) (1+ \bm{\sigma}_2 \cdot \bm{\sigma}_i)
 \end{eqnarray}
where fields are in units of $\Delta_{hf}$ (the width of the Gaussian distribution of hyperfine fields), energies in units of $E_{hf}$, and rates in units of $\omega_{hf}$.
The first two Hamiltonians yield conventional OMAR as described earlier. The last Hamiltonian describes the galvinoxyl spins interacting with the MEH-PPV bottleneck spins via exchange.

First, consider the dilute limit where the distance between the nearest dopant spin to the bottleneck is greater than a critical length, $r_{crit}$. The critical length is defined as the length below which the exchange strength is greater than the hyperfine strength.
Assuming an exchange coupling, $J(r) = J_0 e^{-2 r /\ell}$ ($J_0$ is the on-site exchange constant and $\ell$ is the localization length), the critical length is $r_{crit} = -\frac{\ell}{2}\text{ln}(E_{hf}/J_0)$.
As the exchange varies so strongly with distance, $J(r_1) \gg J(r_2)$, if we define $r_1$ to be the distance between the dopant spin and nearest of the two bottleneck sites (so $r_1 < r_2$ necessarily).
So the dopant spin can be assumed to be just interacting with site 1; the dopant spin adds another source of spin-mixing in addition to the hyperfine fields at the bottlenecked sites.
This new source of spin-mixing reduces the efficacy of the applied field in suppressing singlet-triplet transitions and therefore the magneto-response is lessened as shown in the ``dilute" regime of Fig.~\ref{fig:regimes}.

The ``semi-dilute" regime of Fig.~\ref{fig:regimes} is reached when the dopant is within $r_{crit}$ of \emph{one and only one} of the bottleneck components.
The probability for the bottlenecked spins, in the slow hopping limit, to be in either the singlet ($S = |\uparrow \downarrow \rangle - |\downarrow \uparrow \rangle$) or triplet ($T_0 = |\uparrow \downarrow \rangle + |\downarrow \uparrow \rangle$, $T_{\uparrow} = |\uparrow \uparrow \rangle$, $T_{\downarrow} = |\downarrow \downarrow \rangle$) states can be determined by solving for the density matrix at long times.\cite{Harmon:2012:PRB} For example, in high fields the probability is equally split between the S and T$_0$ states.
However the existence of the third spin modifies these probabilities and reintroduces $T_{\uparrow}$ and $T_{\downarrow}$ components.
The probability of a S$\rightarrow$S transition can be determined by\cite{Harmon2012c}
\begin{equation}\label{eq:ps}
p_{S\rightarrow S} =
  \frac{1}{2}\sum_{m = 1}^8 | \langle m | P_S | m \rangle|^2,
\end{equation}
where $m$ labels the eigenvectors of the relevant spin Hamiltonian and $P_S$ is the singlet projection operator.
The large exchange coupling modifies the probability to $p_{S\rightarrow S} (B_0 \rightarrow \infty)=3/8$.
The zero-field probability changes from $1/3$ to $7/24$.
The difference between no exchange and large exchange is that $p_{S\rightarrow S} (B_0 \rightarrow \infty) - p_{S\rightarrow S} (B_0 \rightarrow 0)  $ is exactly twice as small in the latter case which is precisely the amount $|$MC$|$ decreases by at the lower temperatures (that is, the plateau is half the height of the MC at $f = 0$).
Aside from this mathematical argument for the plateau, a more physical argument can also be supplied.
Again assuming high applied field, consider the state $|S;\uparrow\rangle$ ($S$ represents the spin configuration of the bottleneck and $\uparrow$ indexes the spin of the galvinoxyl radical) which at any moment has a 1/4 chance of being the three-spin configuration.
Fast exchange induces spin flip-flops between site $1$ (which is in S formation with site $2$) and site $g$ which mixes in $T_0$ and $T_{\uparrow}$ states as made explicit in row three of Table I.
The probability to be $|S;\uparrow\rangle$ is now 1/8 because the same probability is assigned to the flip-flopped state; the probabilities of the $T_0$ and $T_{\uparrow}$ can also be resolved as described in the caption of Table I, yielding what we obtained using Eq. (\ref{eq:ps}).

Finally as the doping is increased further, the ``quenching" regime is reached where the MC decreases to zero. With the increase in dopants, the chance increases for a galvonxyl spin to be within $r_{crit}$ of \emph{both} bottleneck sites. It is known that strong exchange between the two bottleneck sites quenches OMAR because spin flip-flops occur faster than the hyperfine fields can operate on the spins;\cite{Wagemans2011, Harmon2012c} thus the spin configuration is effectively pinned in their initial state.
A similar phenomenon occurs in the doped system under consideration here; the galvinoxyl acts as an intermediary between the bottleneck spins, through which they communicate, and ceases any spin evolution.

Quantitative support for the above arguments is provided by
solving  a bipolaron model with the stochastic Liouville equation\cite{Kubo1963, Haberkorn1976,supp}. This method has been a popular choice for determining organic MC due to its flexibility in treating a variety of reactants.\cite{Bobbert:2007:PRL, Schellekens2011}
\begin{table}[h]
\large
\begin{tabular}{| c || c | c | c | c | c |}
\multicolumn{5}{c}{}\\
\hline
$\frac{1}{8}$&  &$\frac{1}{8}\times \frac{1}{4}$ & $\frac{1}{8}\times \frac{1}{4}$&$\frac{1}{8}\times \frac{1}{2}$& $\frac{1}{8}\times \frac{1}{2}$ \\ \hline\hline
\small{$|\xi_0 \rangle$}	& \small{$|\xi_e \rangle = $} & $|a \rangle + $ &$|b \rangle + $	      & $|c \rangle + $  & $|d\rangle$ \\
\hline
\small{$|S; \uparrow \rangle$}	& \small{$1 \stackrel{exch.}{\leftrightarrow} g$} & \small{$\frac{1}{2} |S; \uparrow \rangle$} & \small{$\frac{1}{2} |T_0; \uparrow \rangle$}	      & \small{$-\frac{1}{\sqrt{2}} |T_{\uparrow}; \downarrow \rangle$ } & $0$ \\
\hline
\small{$|S; \downarrow \rangle$}	& \small{$1 \stackrel{exch.}{\leftrightarrow} g$} & \small{$\frac{1}{2} |S; \downarrow \rangle$} & \small{$-\frac{1}{2} |T_0; \downarrow \rangle$} & $0$	      & \small{$\frac{1}{\sqrt{2}} |T_{\downarrow}; \uparrow \rangle$ }  \\
\hline
\small{$|T_0; \uparrow \rangle$} & \small{$1 \stackrel{exch.}{\leftrightarrow} g$} & \small{$\frac{1}{2} |S; \uparrow \rangle$} &\small{$\frac{1}{2} |T_0; \uparrow \rangle$}	      & \small{$\frac{1}{\sqrt{2}} |T_{\uparrow}; \downarrow \rangle$}  & $0$ \\
\hline
\small{$|T_0; \downarrow \rangle$}	& \small{$1 \stackrel{exch.}{\leftrightarrow} g$} &  \small{$-\frac{1}{2} |S; \uparrow \rangle$} & \small{$\frac{1}{2} |T_0; \uparrow \rangle$}	& $0$      & \small{ $\frac{1}{\sqrt{2}} |T_{\downarrow}; \uparrow \rangle$}   \\
\hline
\end{tabular}
\caption[]{Left column: the four possible spin states in high magnetic field.
Right section: spin at site one and galvinoxyl spin-exchanged spin states.
Top row: time-averaged weighting; each element has that probability of occurring in the long time limit when the bottlenecked carrier finally hops. For example, the probability of a singlet is $\frac{1}{8}+\frac{1}{8}$ from the uncoupled spin state options and $\frac{1}{32}+\frac{1}{32} + \frac{1}{32}+\frac{1}{32}$ which sums to a total of $\frac{3}{8}$.}
\end{table}
If we take the dopant spin to interact with the nearest of either site 1 or site 2, the same behavior seen in regions I (dilute) and II (semi-dilute) of Fig.~\ref{fig:regimes} emerges except at higher doping fractions (not shown).
Such analysis assumes the galvinoxyl placement to be independent of bottlenecks.
However better results are found if this assumption is not made;
we discover good agreement with the experimental features for regions I and II (solid red curve in Fig.~\ref{fig:regimes}) when considering the dopant spin interacting solely with site 1 across a distance specified by the probability distribution function\cite{Hertz1909, Torquato1995}
$p_3(r, 1) = 3 \lambda r^{2} \text{exp}({-\lambda r^3})$,
where $\lambda = 4 n_g \pi/3$. \cite{supp}
This suggests the dopant interacts preferentially with site 1 instead of site 2.
It is reasonable to think of the bottlenecks forming between two non-identical sites | for instance site 1 might be a deep trap.\cite{Rybicki2012, Harmon2014}
Likewise, the galvinoxyl dopant may find it energetically favorable to lie near one site over the other.
As the temperature is raised, this galvinoxyl finds more freedom and hence is able to interact better with site 2. The result is a degradation of the plateau (both its length and ratio compared to $f=0$) as observed in Fig.~\ref{fig:concentration}.

At some fraction, the dopants are sufficiently numerous that a significant amount can interact strongly with both MEH-PPV sites (e.g. dopants that fall in the overlap shown in Fig.~\ref{fig:regimes}); this is the quenching regime.
It can be modeled (orange dotted line in Fig.~\ref{fig:regimes}) by choosing the dopant nearest to the midpoint between sites 1 and 2, given suitably large $r_{crit}/a$.
It should be noted that this criterion does not adequately account for the lower doping regimes, so we have plotted the quenching regime shifted to the doping amount where it appears to dominate.

Our results are consistent with the non-monotonic behavior seen in the efficiency versus doping fraction in  P3HT/PCBM solar cells.\cite{Zhang2012, Zhang2013}
Low and high doped solar cells displayed roughly the same efficiency marked by a substantial increase as intermediate fractions.
We understand these results qualitatively in the following manner: at low dopings, the galvinoxyl interacts solely with the PCBM electron which increases the likelihood of inert triplets thereby enhancing efficiency. Increasing the doping by too much though allows the galvinoxyl to strongly interact with the PCBM electron and the P3HT hole which pins the spin states to what they are at the time of their encounter (i.e. as if there were no doping).
The values used in our calculations can be found in the Supplementary Material.\cite{supp}

In conclusion, we have provided the first magneto-transport measurements in radical-doped organic semiconductors.
A feature where the saturated magneto-conductance plateaus over a wide range of doping fraction is observed over all temperatures investigated.
This phenomenon is well-explained by a theory in which a single dopant spin strongly interacts, by exchange, with \emph{one} of the bottleneck sites.
We attribute this effect also to the efficiency increases observed in organic solar cells for certain doping fractions.

This work was supported by Army MURI Grant No. W911NF-08-1-0317.

\end{document}